\documentclass[12pt, preprint]{aastex}

\shorttitle{Earth's spectral apparent albedo}
\shortauthors{Monta\~n\'es-Rodr{{i}}guez et al.}

\begin{document}

%% LaTeX will automatically break titles if they run longer than
%% one line. However, you may use \\ to force a line break if
%% you desire.

\title{Globally integrated measurements of the Earth's visible
spectral albedo}

%% Use \author, \affil, and the \and command to format
%% author and affiliation information.
%% Note that \email has replaced the old \authoremail command
%% from AASTeX v4.0. You can use \email to mark an email address
%% anywhere in the paper, not just in the front matter.
%% As in the title, use \\ to force line breaks.

\author{P. Monta\~n\'es-Rodr{{i}}guez, E. Pall\'e and P.R. Goode}
\affil{Big Bear Solar Observatory, New Jersey Institute of Technology,
Newark, NJ 07102, USA}
\email{pmr@bbso.njit.edu}

\author{J. Hickey}
\affil{Palomar Observatory, 35899 Canfield Rd., Palomar Mountain,
CA 92060-0200, USA}

\and

\author{S.E. Koonin}
\affil{W.K. Kellogg  Radiation Laboratory, California Institute of Technology,
Pasadena, CA 91125, USA}

\begin{abstract}
We report spectroscopic observations of the earthshine reflected
from the Moon.  By applying our well-developed photometry
methodology to spectroscopy, we were able to precisely determine the
Earth's reflectance, and its variation as a function of wavelength
through a single night as the Earth rotates. These data imply that
planned regular monitoring of earthshine spectra will yield
valuable, new inputs for climate models, which would be
complementary to those from the more standard broadband measurements
of satellite platforms. For our single night of reported
observations, we find that Earth's albedo decreases sharply with
wavelength from 500 to 600\,nm, while being almost flat from 600 to
900\,nm. The mean spectroscopic albedo over the visible is
consistent with simultaneous broadband photometric measurements.
Unlike previous reports, we find no evidence for an appreciable
``red'' or ``vegetation edge'' in the Earth's spectral albedo, and
no evidence for changes in this spectral region (700\,-740\,nm) over
the 40\,$^\circ$ of Earth's rotation covered by our observations.
Whether or not the absence of a vegetation signature in
disk-integrated observations of the earth is a common feature awaits
the analysis of more earthshine data and simultaneous satellite cloud 
maps at several seasons. If our result is confirmed, it would limit 
efforts to use the red-edge as a probe for Earth-like, extra-solar
planets. Water vapor and molecular oxygen signals in the visible
earthshine, and carbon dioxide and methane in the near-infrared, are
more likely to be a powerful probe.

\end{abstract}


\setcounter{equation}{0}
\section{Introduction}

Ground-based measurements of the earthshine (ES) have become a
valuable tool for studying Earth's global climate in which one
views the Earth much as though it were another planet in the solar
system. A precise determination of changes in Earth's total or
Bond albedo is essential to understanding Earth's energy balance.
While measuring the reflectivity for other planets is a direct
task, the indirect observation of the earthshine, after its
reflection from the Moon, is the only ground-based technique that
allows us to measure the albedo of our own planet.

In previous publications, we have used long-term photometric
observations of the earthshine to determine variations in Earth's
reflectance from 1998 to 2003 (Goode et al., 2001; Qiu et al.,
2003; Pall\'e et al., 2003), and by comparison with International
Satellite Cloud Climatology Project (ISCCP) datasets, we
reconstructed the Earth's global reflectance variations over
the past two decades (Pall\'e et al., 2004). Those variations in albedo
depend on changes in global cloud amount, cloud optical thickness and
surface reflectance of the earthshine-contributing parts of Earth.

Here we demonstrate the potential of spectroscopic studies of the
earthshine in climate studies.  We do this by quantifying the
wavelength dependence of Earth's albedo through a single night of
observations. A determination of the longer-term evolution of the
spectrum awaits more spectral data. Secondarily, our spectral
study enables us to measure the strength of the ``red'' or
``vegetation edge'' in the earthshine. Knowing the strength of
this latter signal is critical in evaluating the utility of
exploiting the spectral ''vegetation edge''  in the search for
Earth-like extra-solar planets. Previous efforts to estimate the
globally averaged spectrum of Earth (Arnold et al., 2002; Woolf et
al., 2002) have aimed to give a qualitative description of its
astrobiological interest.

We measured the albedo on 2003/11/19 over the sunlit part of the
Earth, centered for that night over the Atlantic Ocean, by doing
spectrophotometry ($R=\lambda/\Delta\lambda\sim 1000$) of the
earthshine (ES, reflected from the dark side of the Moon) and
moonshine (MS, the bright side of the Moon) from Palomar
Observatory. The appropriate lunar geometry corrections were
applied to determine a mean spectral dependence comparable to
photometric albedos simultaneously taken at Big Bear Solar
Observatory.   Synthetic simulations of the photometric albedo
(Pall\'e et al, 2003) were also made and compared favorably.

In Section 2 of this paper, we first describe the methodology used
to determine the wavelength-dependent apparent (nightly) albedo
from earthshine observations. In Section 3, we give details about
our data acquisition process, data reduction and analysis. In
Section 4, we describe how the relative lunar reflectivity for
each of the observed patches was determined. In Section 5, we
present our results for the evolution of the spectral albedo
through the night of observations and describe the significant
spectral features. Comparisons with models and observations of
photometric albedo taken during a consecutive night are also
included in that section. We discuss the implications of the
results in Section 6.

\section{Earth's wavelength-dependent apparent albedo}

Qiu et al. (2003) developed the methodology that provides the
basis used here to calculate the apparent albedo from earthshine and
moonshine measurements. We briefly extend that description by
introducing the wavelength dependence. An averaged spectral Bond albedo
can be determined by means of continuous observations of earthshine by
integrating over all lunar phases.

In this study, we measure the Earth's apparent albedo, $p^*(\beta,
\lambda)$ at lunar phase angle $\theta=$ +117$^\circ$ (one can see
from Figure 1 of Qiu et al. that $\theta$ and $\beta$, the Earth's
phase angle, are essentially supplementary angles). The apparent
albedo for a given night is described by the Equation (Equation
(17) of Qiu et al.),
\begin{equation} p^*(\beta, \lambda) =
\frac{3}{2f_L}\frac{p_{b,\lambda
}f_b(\theta)}{p_{a,\lambda}f_a(\theta_0)}
\frac{I_{a,\lambda}/T_{a,\lambda}}{I_{b,\lambda}/T_{b,\lambda}}\frac{R^2_{EM}}{R
^2_{E}}\frac{R^2_{ES}}{R^2_{MS}},
\label{p^*} \end{equation} where $a$ and $b$ represent the
observed patches on the earthshine and moonshine sides of the
Moon,  respectively. $I_{a,\lambda}$ is the earthshine radiance as
observed from the ground and  $T_{a,\lambda}$ is the local
atmospheric transmission, thus the ratio $I_{a,\lambda}/T_
{a,\lambda}\over I_{b,\lambda}/T_{b,\lambda}$ gives us the
 earthshine intensity relative to the moonshine intensity atop the
terrestrial  atmosphere. ${p_b\over p_a}$ and ${f_b\over f_a}$ are
the ratios of the patches geometrical albedos and lunar phase
functions, respectively.  $f_L$ is Earth's Lambert phase function,
while $R_{EM}$, $R_{MS}$, and $R_{ES}$ are the Earth-Moon,
Moon-Sun, Earth-Sun distances and $R_E$ is the Earth's radius. We
expect any wavelength dependence in the lunar phase function to be
canceled in the ratio, ${f_b\over f_a}$.  Note that $\theta_0$ is
the angle subtended between the incident and the reflected
earthshine from the moon, and is of the order of $\le$ 1$^\circ$.

An additional observational correction was applied to account for
the {\it opposition effect} caused by the coherent backscatter of
the lunar soil and the Earth's shadow hiding (Hapke et al, 1998;
Hapke et al, 1993; Helfenstein et al, 1997). This actually
represents a re-calibration in the normalization of our lunar
phase function, since it is significant only at small
($<$5$^{\circ}$) lunar phases. The opposition effect was corrected
as described in Qiu et al., (2003) and was also considered to be
wavelength-independent.

\section{Data acquisition and analysis}

\begin{figure}
\epsscale{0.70}
\plotone{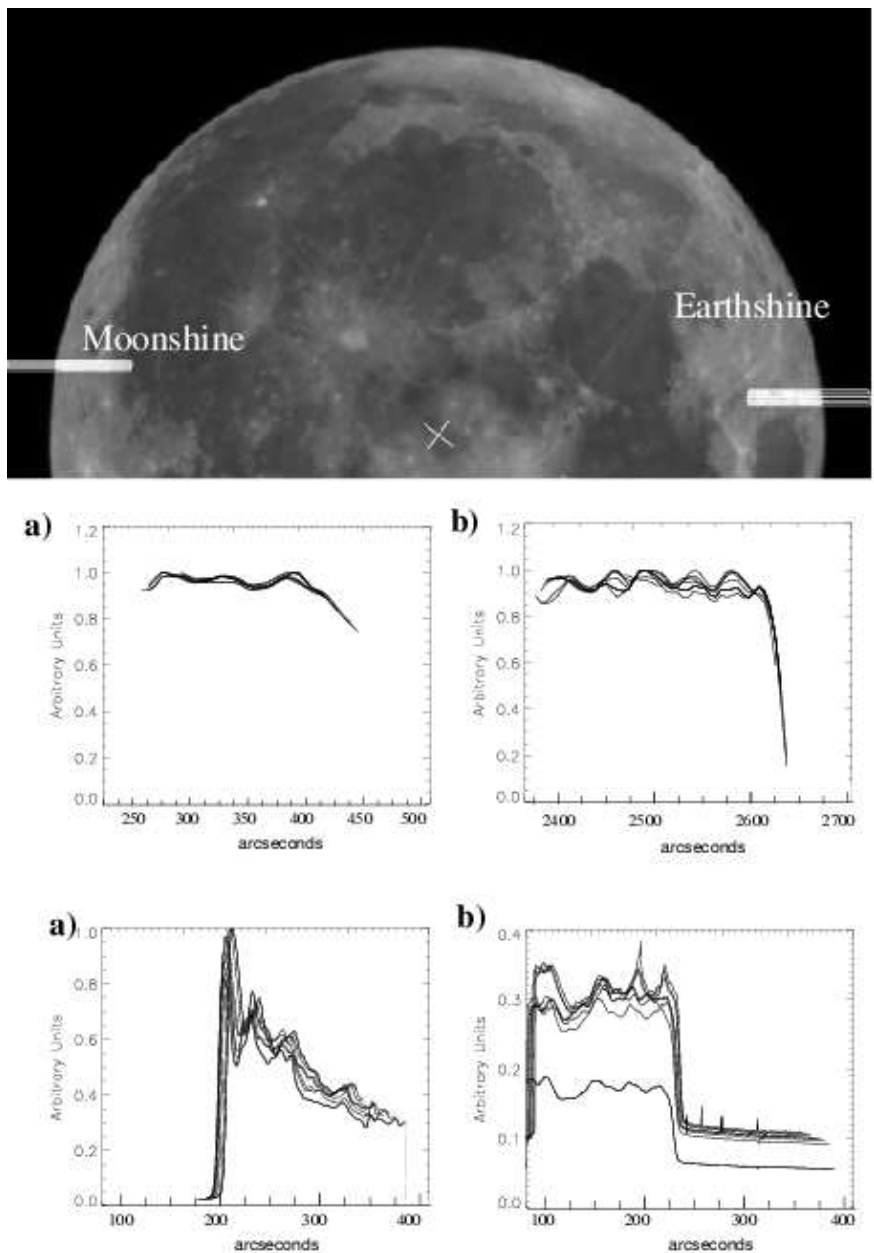}
\caption{Top: Lunar image taken during the 1993/11/29 eclipse. The MS and ES
slit positions during spectroscopic observations on 2003/11/19 are shown for
each MS-ES cycle (best visible as eight individual fine white lines just off
the limb). Middle: MS (a) and ES (b) limb profiles at lunar phase angle =
+1.23$^{\circ}$, (during eclipse observations) for the slit positions on
2003/11/19. Bottom: (a) MS (showing limb brightening) and (b) ES (showing
earthshine flatness on the disk) as actual image profiles along the slit during
spectroscopic observations on 2003/11/19 (note that an offset has been
introduced in the x-axis). The sharp drops in the middle and bottom occur at the lunar limb. The lower profile at the bottom panel (b) is due to the shorter integration time used during the first exposure.}
\label{profiles}
\end{figure}

We observed the spectra of the ES and MS for 2003/11/19, a clear
night over Palomar Observatory, between 10.47 to 13.08 hours (UT).
We observed with a single order, long slit spectrograph using the
Palomar 60$''$ (1.5\,m) telescope. An entrance slit of
1.32 arcsec\,x\,6 arcmin on the sky was selected;
for comparison, the mean lunar apparent diameter
during the time of observations was 32.24 arcmin. This configuration covers the
spectral region between 460\,nm and 1\,040\,nm. However this range was limited
by the poor CCD quantum efficiency below 500\,nm and by the fringe interference patterns in the red, which start to appear at about 800\,nm (although they were reduced when taking the ratio ES to MS) but are always important above 980\,nm. Thus, the final effective range was limited to between 480\,nm and 980\,nm.

The slit was EW oriented and positioned in such a way that about
half of it was over the lunar disk near the limb, and the other
half over the adjacent sky, see Figure~\ref{profiles}. Two lunar patches were
selected near the bright and dark lunar limbs in the highland regions of the moon. We alternatively measured these patches as the airmass varied through the night. During each MS-ES cycle, the moon was tracked in right ascension and declination at a fixed rate. To account for the moon's motion relative to the stars, rates were updated every 15 minutes, to minimize the deviation from the original patch.  The updates were done after each MS-ES cycle was completed. We obtained a total of 8 MS-ES cycles during observations from moonrise to sunrise.

The basic reduction steps were performed with the Image Reduction
and Analysis facility (IRAF) software system. The alignment between the dispersion direction and the CCD rows was verified and the necessary geometrical corrections were applied to all images in order to preserve the maximal possible spectral resolution during the extraction of the spectra; laboratory Argon lamps were used for this purpose, as well as for wavelength
calibration.

The spatial profiles within the region covered by the slit for
each MS and ES exposure are shown in Figure~\ref{profiles}. These
profiles are affected by three correctable and/or ignorable types
of variations: i) a large gradient due to the overall Lambert-like
illumination of the MS, ii) light scattered from the moonshine
into the ES (which decreases linearly with the distance to the MS
limb), and iii) small features due to the varying reflection index
over the lunar surface within the region covered by the slit.

Eight apertures of equal size were initially extracted from every
image, four over the sky side and four over the lunar side. The
background due to the scattered light was corrected by subtracting
a linear extrapolation of the sky spectrum to the position of each
of the four ES and MS lunar apertures. The extrapolation was done
point by point across the spectral direction. Then, all spectra
were normalized to a unit exposure time.

Because moonshine and earthshine exposures were not exactly
simultaneous, each MS spectral transmission was interpolated to
the airmass of its temporally closer ES spectrum. The ratio ES/MS,
$I_{a,\lambda}/T_ {a,\lambda}\over I_{b,\lambda}/T_{b,\lambda}$ in
Equation~\ref{p^*}, could then be calculated to obtain the implied top of the atmosphere ratio.

\section{Measuring the lunar relative reflectivity}

\begin{figure}[h]
\epsscale{0.99}
\plotone{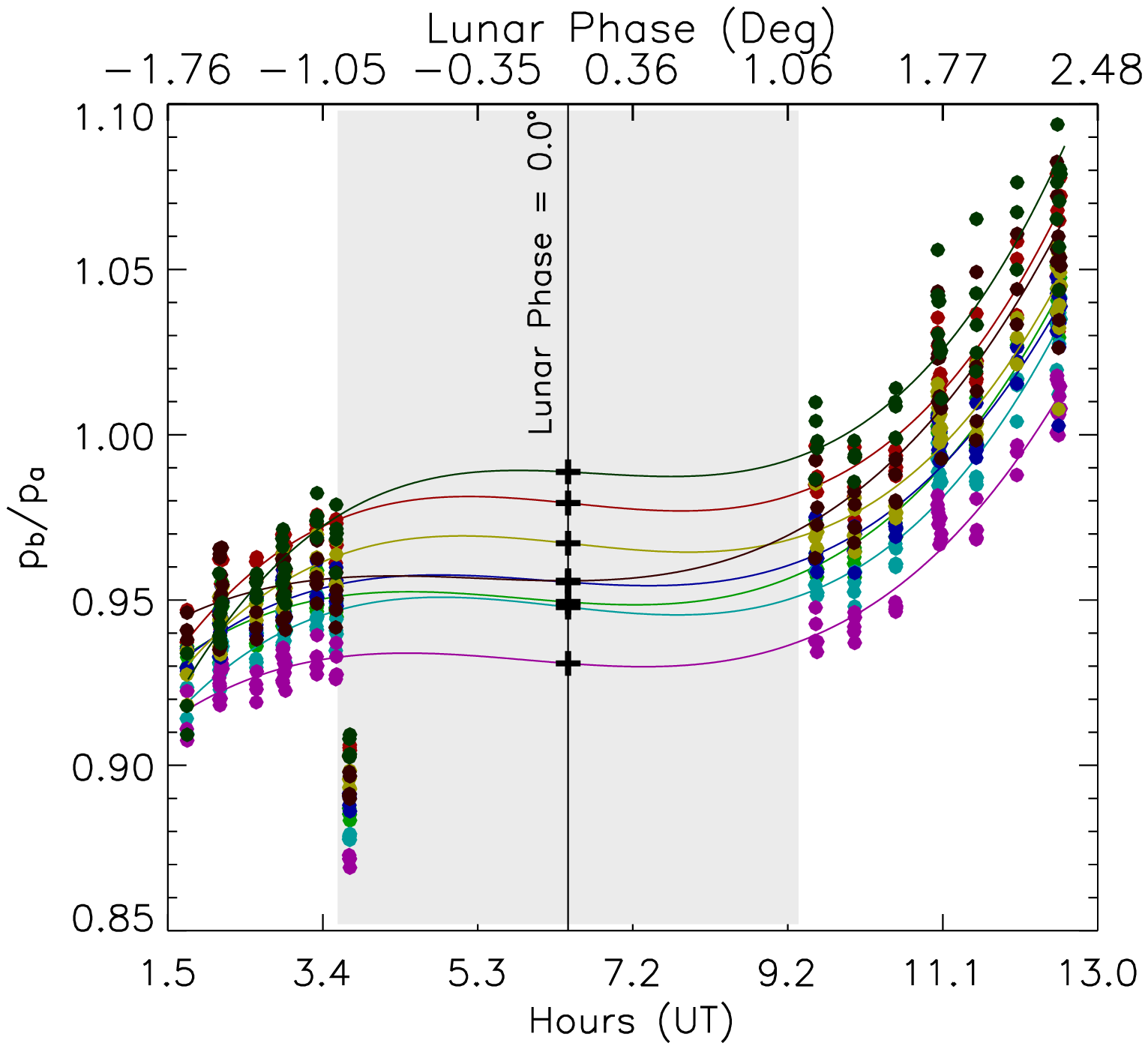}
\caption{Variation of $\frac{p_b f_b}{p_a f_a}$ for all 8 lunar patches
(different colors) during 1993/11/29 eclipse. The ratio only takes
its textbook meaning when the lunar phase angle is zero. The
penumbral shadow region is shaded in the figure, and the best
polynomial fit for each patch is also shown. The ratios at lunar
phase equal zero interpolated from each fit are given in Table 1.
Penumbral shadowing causes one set of points to be well-below the
fitted lines.}
\label{eclipse}
\end{figure}

\begin{table}
\caption{Measured relative reflectivity $p_b/p_a$ at eclipse
totality (lunar phase angle ($\theta$) zero). A standard deviation of
about 0.007 resulting from a covariance calculation of the polynomial
fit was estimated for each point.
The time (UT) at which each ES spectrum was taken and the patch center
in selenographic coordinates are also given.}
\begin{tabular}{cccc}
\hline
Time (UT)  &  $lat_b,\,lon_b$   &$lat_a,\,lon_a$  & $p_b/p_a$ \\
\hline
10.47      & -13.3, -82.4       &  30.2,73.1 & 0.979  \\
10.89      & -12.8, -80.0       &  31.0,73.7 & 0.949  \\
11.26      & -12.7, -81.5       &  31.3,71.1 & 0.947  \\
11.63      & -13.6, -79.5       &  31.9,71.5 & 0.930  \\
12.02      & -13.4, -80.9       &  31.0,73.7 & 0.955  \\
12.38      & -13.8, -82.0       &  30.7,73.5 & 0.967  \\
12.74      & -13.2, -78.6       &  31.5,86.3 & 0.955  \\
13.08      & -13.4, -83.9       &  31.0,86.3 & 0.988  \\
\hline
\end{tabular}
\label{lunar_ref}
\end{table}

To determine the Earth's reflectance on 2003/11/19, one requires
two quantities from observations on other nights -- the Moon's
geometrical albedo (determined from total eclipse data) and the
lunar phase function (determined over several years of photometric
observations, Qiu et al., 2003). To apply these to each of the
eight pairs of measurements, one requires the accurate
selenographic coordinates for the regions from where the
earthshine and moonshine are reflected (Qiu et al., 2003).

\begin{figure*}[t]
\noindent\includegraphics[width=40pc]{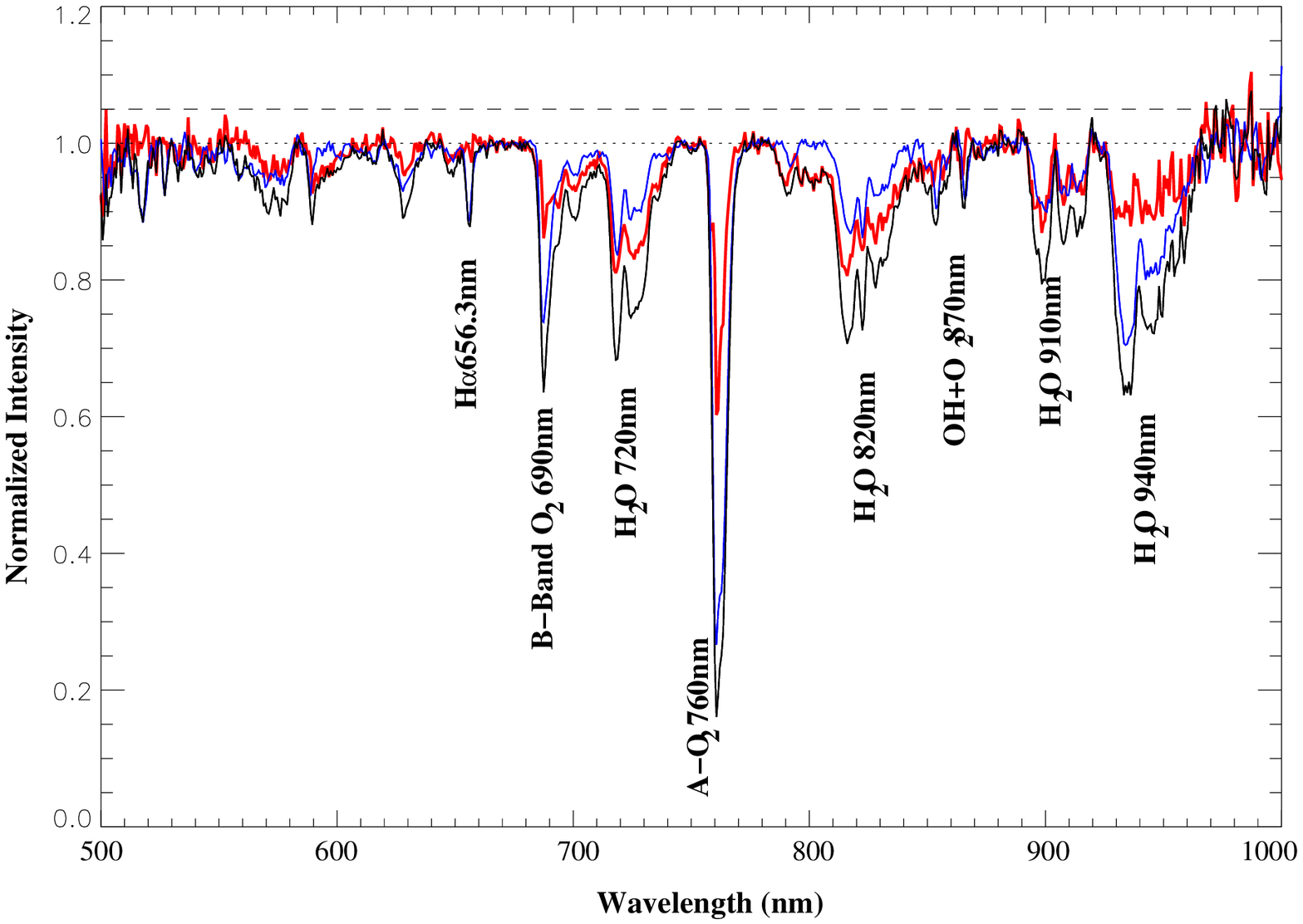}
\caption{Normalized moonshine spectra (blue), earthshine (black)
and their ratio (red) for 2003/11/19. The main gas absorption
bands, as well as other features and their central wavelengths are
indicated. Note how the solar H$\alpha$ line disappears in the ES/MS ratio.}
\label{normal}\end{figure*}

For each of the eight MS-ES spectral observing pairs, we must
re-point the telescope and in the subsequent analysis make a
precise, {\it post facto} determination of the location of the
narrow slit for each observation (rather than always looking at
the same fiducial patch, as we do in the photometric
observations). We estimate a deviation of up to 10 arcsec in the
EW direction and up to 20 arcsec in the NS direction. This
assessment is consistent with changes in the slit position shown
in Figure~\ref{profiles}, (top panel) where each narrow white line
(best viewed just off-the-limb) represents the location of
spectrograph's slit for each observation. The magnitude of the
changes in the lunar profile observed for different exposures are
shown in the same figure for lunar phase +1.23$^{\circ}$ (middle
panel) and lunar phase +117$^{\circ}$ (bottom panel).

Qiu et al. (2003) argued that the lunar geometrical albedo for the
earthshine and moonshine differ by only a small spectral offset,
equivalently, we can assume here that $p_{b,\lambda}/p_{a,\lambda}=p_{b}/p_{a}$.
In the data reduction, the relative photometric reflectivity of the patches $b$
and $a$ was determined for each of the eight pairs of MS and ES images via a {\it post facto} approach to precisely eliminate the error
introduced by small deviations from the initial selenographic
coordinates (fiducial patch) when re-pointing the telescope for
each MS-ES observing cycle.  To measure each ratio, $p_{b}/p_{a}$,
we used photometric observations of the full moon, interpolating
at lunar phase angle $= 0.0^{\circ}$ during the  lunar eclipse on
1993/11/29.  This eclipse was observed during eleven hours (before
and after totality) from Big Bear Solar Observatory (BBSO) using
the 25\,cm solar telescope that is now part of BBSO's global
H$\alpha$ network. Between four and five images were taken,
approximately, every 30 minutes. Sixteen lunar patches of 3 arcmin
(EW) x 2 arcsec (NS) centered on each ES and MS slit position were
selected. The different Earth-Moon distance  and lunar polar axis
orientation, for the eclipse night and for observations on
1993/11/29 were taken into account. The differences in lunar
libration were considered by properly modifying the projected size
of the observed patches for each eclipse image. Intensities were
normalized to one second exposure and the mean reflectivity for
each patch was then calculated to obtain a pair of $p_b$ and $p_a$
values for each MS-ES observed cycle.
Figure~\ref{eclipse} shows the variation of the ratio
$\frac{p_b f_b(\theta_b)}{p_a f_a(\theta_a)}$ for each
of the eight location pairs corresponding to the eight pairs of
observations on 2003/11/19. The location of the pairs and the
extrapolated ratios are shown in Table 1. To extrapolate our
``$p_b/p_a$'' (quotation marks serve to remind that the ratio is
actually defined only for lunar phase angle zero degrees)
observations to zero phase angle, a polynomial fit was done for
all ratio values just before and after the moon was within the
Earth's penumbral shadow. Note
that one set of points does not come close to the curves because
this set was taken when the Moon was partly in Earth's shadow.

For the lunar phase function, we use the values of Qiu et al.
(2003) after evaluating the phase function for the bands in
Figure~\ref{profiles} and finding that the phase functions are all
essentially the same, after removing the ratio $p_b/p_a$ that
provides the overall normalization.

\begin{figure*}
\noindent\includegraphics[width=38pc]{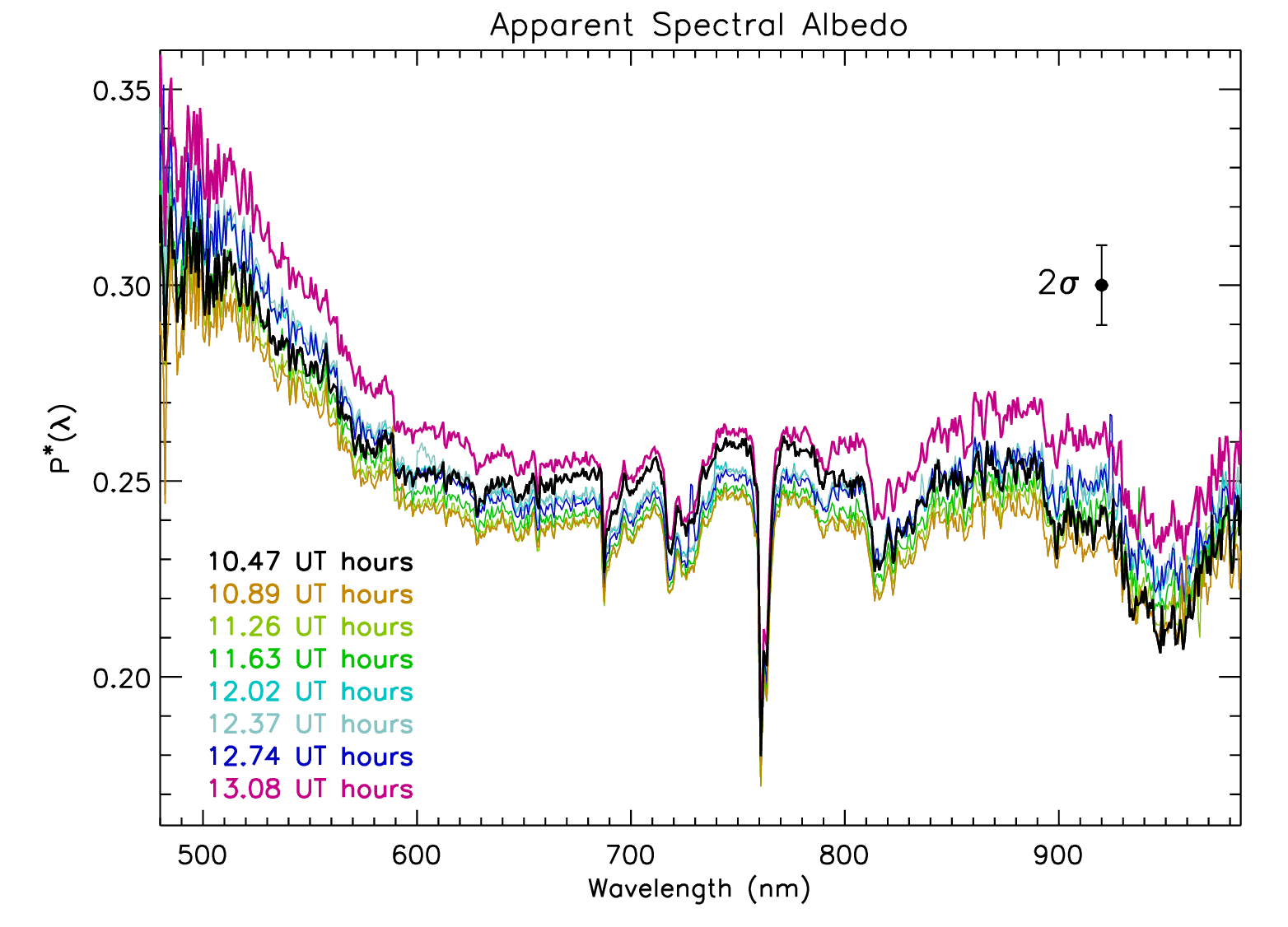}
\caption{Apparent albedo $p^*(\lambda)$ for the night of observations.
The main features of the Earth's reflectance in this region include an
enhancement due to the Rayleigh Scattering in the blue, part of the Chappuis
O$_3$ band, which contribute to the drop above 500\,nm. Atmospheric absorption
bands due to oxygen, as the sharpest A-O$_2$ at 760\,nm, and water vapor are
clearly detected. The surface vegetation edge, which is expected to show an
apparent bump in the visible albedo above 700\,nm is not strong, and neither
does it seem to vary appreciably or systematically through the night.}
\label{pl}
\end{figure*}

\section{Results}
\begin{figure*}
\noindent\includegraphics[width=30pc]{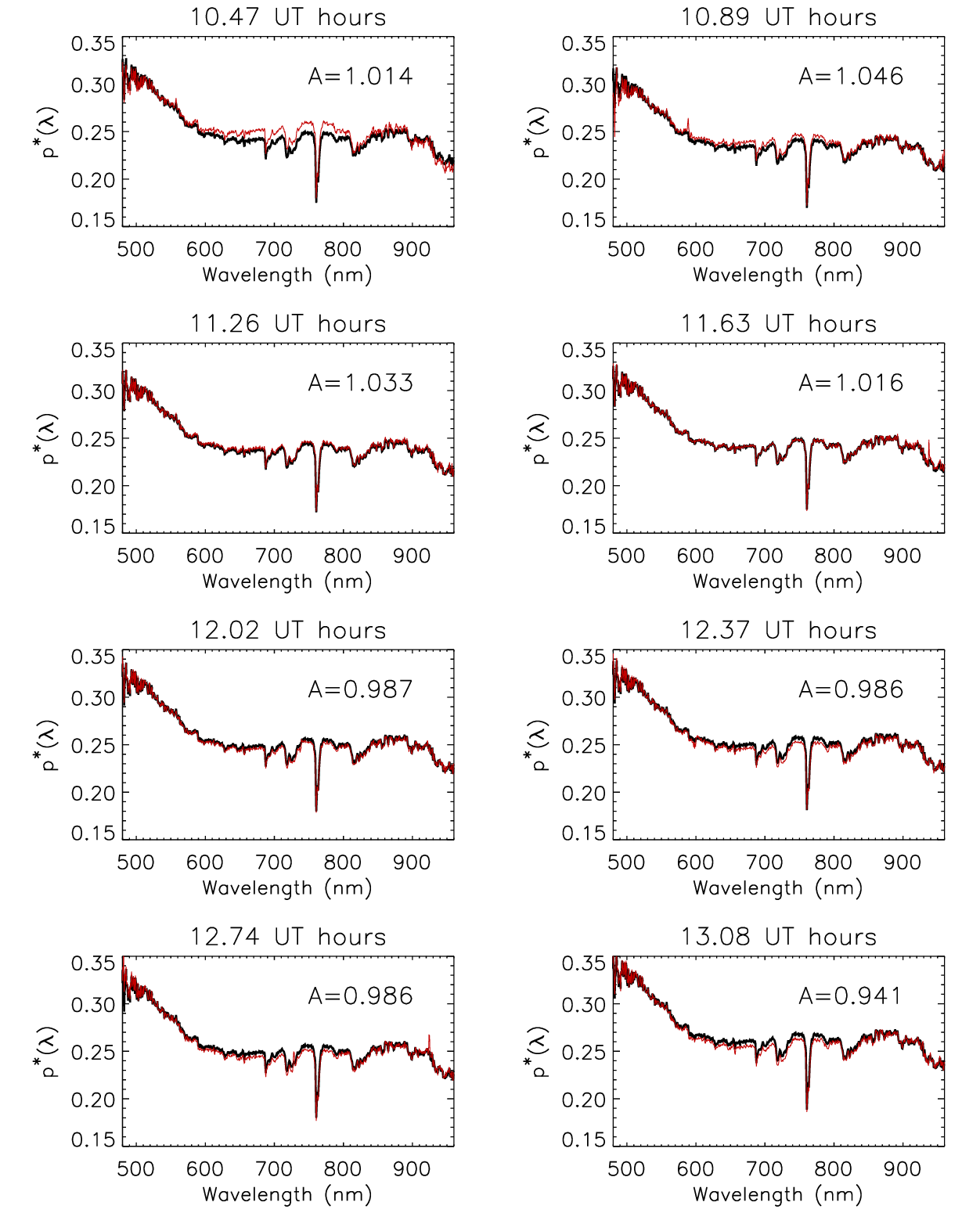}
\caption{Apparent albedo $p^*(\lambda)$ as a function of wavelength
taken during the night of observations (red). Also plotted (black) is
the mean spectrum for the night, $<p(\lambda)>'$, the mathematical
average of our eight $p^*(\lambda)$ spectra. The averaged spectra, p'
has been artificially displaced by a factor A, also given in each panel,
to match the mean value of each $p^*(\lambda)$ over the 500-550\,nm
interval. Note that for the first spectra our observations were centered
over Africa (upper left panel), while for the last ones we were monitoring
the Amazonian Forest (lower right panel). A deviation from the mean value
of up to a 4\% in appreciated in the first measurement.}
\label{evolution}
\end{figure*}

The atmospheric absorption bands in the moonshine spectrum are
formed when the moonlight passes through the local atmosphere. The
bands in the earthshine spectrum additionally contain the
absorption formed when the sunlight passes twice through the
global terrestrial atmosphere, on the sunlit part of the Earth,
before being reflected from the moon toward our telescope
(Monta\~n\'es Rodr{{i}}guez et al., 2004). Thus, absorption lines
in the ES spectrum are deeper than in the MS spectrum. Since the effect
of the local atmosphere is present in both MS and ES, it is removed when
the ratio ES/MS is calculated. In Figure~\ref{normal}, the Earth's
atmospheric spectral features,
primarily due to molecular oxygen and water vapor, are appreciable
for the ES, for the MS and for their ratio. As one would expect,
ES features are deeper than MS features whereas lines formed in
the solar photosphere, such as H$\alpha$ remain invariant for ES
and MS and therefore disappear in the ratio, which further
convincing evidence that the last pass through atmosphere is also
canceled out in the ratio. Only the cloud patterns in the global
atmosphere will affect the shape of the ES/MS spectra by altering
the optical path of the ES depending on cloud altitudes and
optical thicknesses.

\subsection{Nightly spectral albedo} \label{result_intro}

\begin{figure}[h]
\noindent\includegraphics[width=25pc]{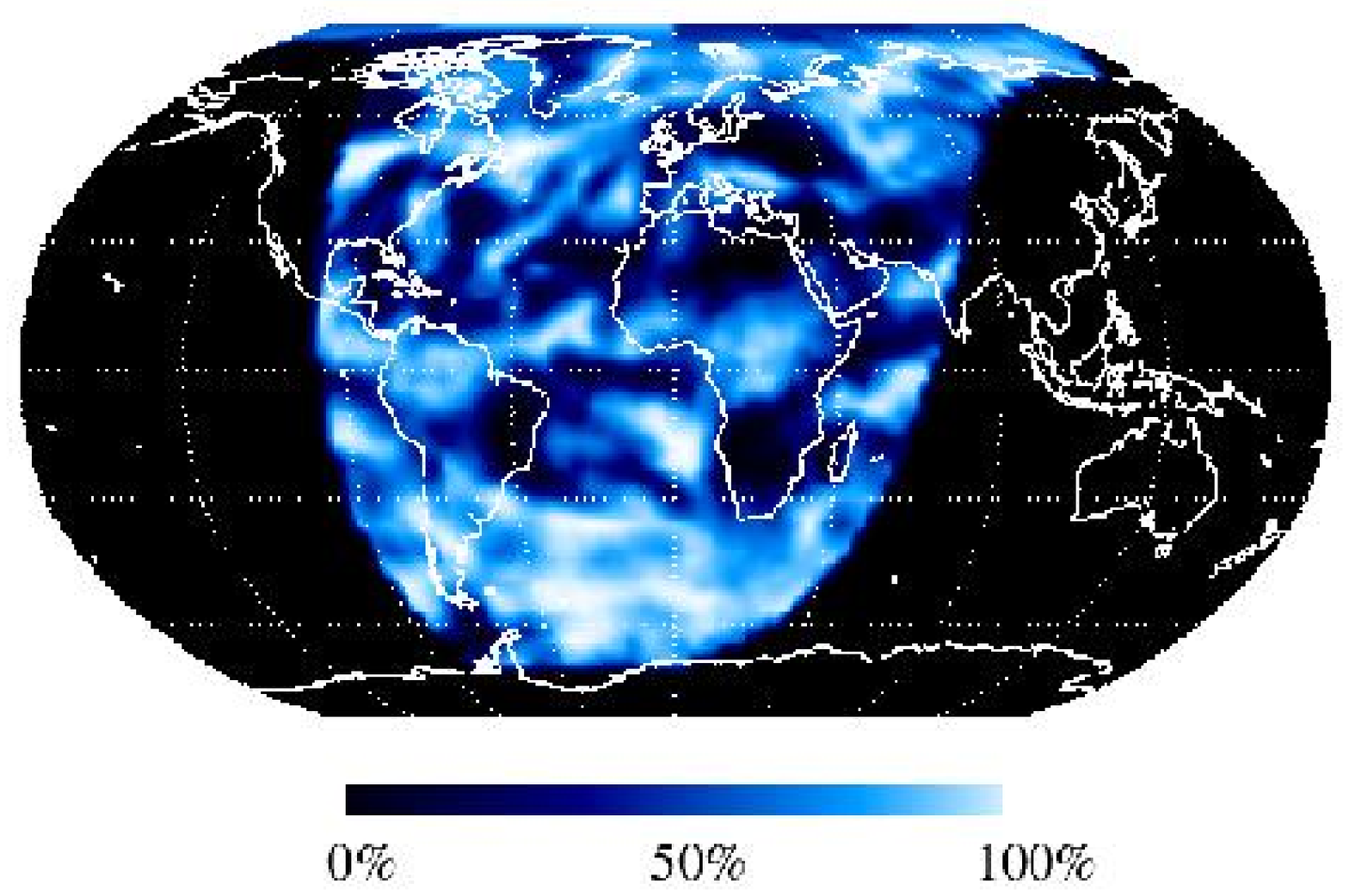}
\caption{Earthshine-contributing area of the Earth during morning
observations from California on 2003/11/19 (10:45-13:10 UT). The
superimposed color map represents the mean cloud amount from ISCCP
data over this area. Since ISCCP data for 2003 are not yet
available, the cloud cover map for 2000/11/19 (closest
``equivalent'' available) is here shown.}
\label{mapa}
\end{figure}

The measured apparent albedos, $p^{*}(\lambda)$, are shown in
Figure~\ref{pl},  where different colors indicate its temporal
evolution through the night,  while different land and cloud
distributions move into the sunlit side of the Earth and are visible
from the Moon.

During our observing night, we were monitoring the Earth's
spectral reflectance over the area shown in Figure~\ref{mapa}.
Because ISCCP data are still not updated up to 2003, cloud data
for 2000/11/19 were used in the figure for illustration. South
America was only partially visible at 10:45 hours, but was near
the center of the sunlit area at 13:10 hours. The total covered
area during the 2.6 hours of observations, without considering
cloud cover, is composed of 79-71\% of oceans, 8-11\% snow  or ice
covered areas, and 13-18\% of land, 9-14\% of which are vegetated
areas mainly located  in the Amazonian rain forest, Equatorial
Africa and Europe. The range of percentages for each surface type
correspond to 10:00 and 13:00 UT, respectively.

The possibility of using the red edge signature of vegetation as a
tool to detect earth-like extra-solar planets has been discussed
in the literature. The red edge has been detected in surface spectral albedo
measurements from airplanes at low altitude
and from the Galileo spacecraft, all these measurements taken with some
spatial resolution. The detected signal shows a step in the Earth's spectrum
starting at about 720\,nm, coinciding with an important water absorption band
in the atmosphere, and continues to the near infrared (Wendish et al., 2004).

Woolf et al. (2002) observed the earthshine an evening in June 2001
(therefore they were monitoring the Pacific Ocean), and they report an
inconclusive vegetation signature above 720\,nm, although at the time of
their observations no large vegetated area of the globe was visible from
the Moon. More detailed observations of the earthshine were undertaken by
Arnold et al (2002). They concluded in their study that the vegetation edge
was difficult to measure in their earthshine data due to clouds and the
effect of atmospheric molecular absorption bands. One of our purposes in
this work was to confirm these results and to report any diurnal
variability in the vegetation edge or other earthshine signatures, if any.

During the night of 2003/11/19, we find no notable red edge
enhancement, even at the times when the South American continent is
in full view and in November, when the ``greenery'' is abundant.
This is not a surprise because we should expect that some, if not
all, of the vegetated areas contributing to the earthshine to be
obscured by the presence of clouds. However, when globally
monitoring the Earth, one should note that at all times about 60\%
of the planet (Rossow et al., 1993) is covered by clouds  with a
mean global cloud amount variation of about 10\% from autumn to
spring.

The possibility remains thus, that the whole or large part
of the South American continent was covered by clouds at the time of
observations, masking any vegetation signal. Mean monthly and annual
cloud amounts from ISCCP indicate that this is frequently the case
throughout the year ({\tt www.isccp.giss.nasa.gov}). This could
easily be solved by analyzing cloud cover maps for the same night of
our observations. If the night of 2003/11/19 had mostly clear skies
over that area, then that would imply that the vegetation signal is
{\it per se} too small to be detected in disk-integrated measurements of
earthshine. On the contrary, if it was mostly cloudy, the signal may
be strong enough to be detected if clear skies occur over the large
enough portion of the rainforests regions of the earth. However, this is
not typically the case, as densely vegetated areas are associated
with vegetation transpiration and enhanced cloud formation (IPCC,
2001).

The enhancement of the reflectivity in the blue parts of the
spectra, below 600\,nm, due to the effect of atmospheric gas
molecules producing Rayleigh scattering is the most significant
feature in our results.  The albedo increase toward the blue is
consistent with other surface spectral albedo measured from
airplanes (Wendisch et al., 2004). From spatially resolved
observations and spectral albedo models, a peak at around 310\,nm
would be expected, if we were covering that spectral region. This is
due to the combination of Rayleigh scattering, which decreases as
$\lambda^{-4}$, and an ozone absorption.

An enhancement in the 600 to 900\,nm region, or else a decrease in
the red and blue extremes, of  about 4\% with respect to its mean
spectral value is found when we compare the first spectrum taken at
10.47 hours with the temporal averaged of the eight spectra
(Figure~\ref{evolution}). As can be seen in the figure, this
relative bump systematically changes its curvature through the night
(unlike for the 500-600 nm regime), thus we cannot attribute it to
any instrumental problem. Since the maximum enhancement is at the
beginning of the night, when the sun is still rising in South
America, and it almost disappear two hours later, we cannot
associate this enhancement with the signal of vegetation neither,
which should behave in the opposite fashion.  Further, no signal of
vegetation is expected below 700 nm.  Thus, we associate this
variation to the evolution of the surface and cloud pattern over the
sunlit Earth during the time of observations.

\subsection{Precision of the results}\label{errors}

The accuracy of each $p^*(\beta,\lambda)$ derived from our
earthshine measurements was determined from Equation~\ref{p^*}, and
depends on the errors in the earthshine to moonshine ratio, the
lunar reflectivity (which accounts for the relative reflectivity
between two fiducial patches) the lunar phase function  (which
considers the geometrical dependence of the lunar reflectivity) and
the opposition effect coefficient.

The measurement of the earthshine to moonshine ratio, after the sky
subtraction and the Beer's law fitting of the bright side spectra,
is the biggest source of error, as occurs in our photometric
observations (see Qiu et al., 2003 for details). We estimate it as
smaller than 2\%. The error derived from the measurement of the
(relative) lunar reflectivity was taken as the standard deviation of
the polynomial fit to $\frac{p_b f_b(\theta_b)}{p_a f_a(\theta_a)}$,
and has a value of 0.7\% from a covariance calculation. The
precision achieved for the lunar phase function (and opposition
effect coefficient) can be determined down to 0.5\%, as report in
section 5 of Qiu et al., 2003.

From the propagation of the errors into Equation~\ref{p^*}, a
standard deviation of the order of 3.7\% is obtained for each
individual spectral albedo measurement. This is shown in
Figure~\ref{pl} by a two sigma error bar. They are also represented
for each data point in Figure~\ref{mean}.

\subsection{Comparison with photometric albedos}\label{evaluation}

\begin{figure}[h]
\noindent\includegraphics[width=25pc]{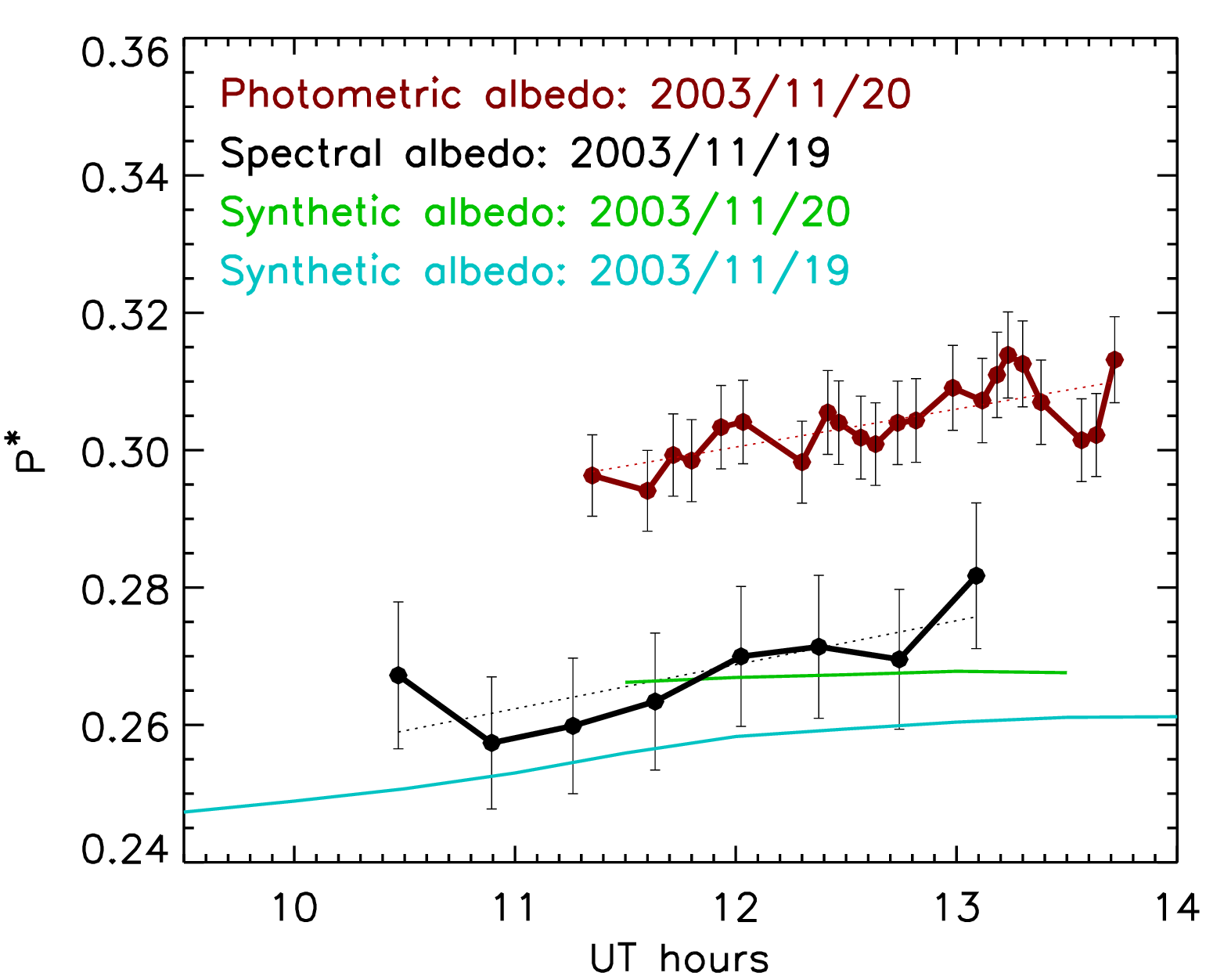} \caption{Temporal
variation of the spectral albedo averaged between 480 and 700\,nm
(black solid line). The error bars show the 2\,$\sigma$
deviation for each spectral measurement determined from the
propagation of errors as discussed in Section 5.2. A linear fit to
all points through the night is also shown. The measured apparent
albedos from photometric observations for a consecutive night are
shown for comparison (red solid line), with error bars also representing
twice the standard deviation achieved, but through photometry. Note that
the spectral coverage for the photometric albedo is broader (between
400 and 700\,nm) than for the spectral albedo.  Also plotted are
the modeled photometric albedos for 2003/11/19 and 2003/11/20
covering the entire short-wave region through the near IR, using
ISCCP data for 2000/11/19 and 2000/11/20 (blue and green). The
differences in wavelength coverages probably accounts for much of
the offsets.} \label{mean}
\end{figure}

An evaluation of our results can be done with a comparison to
photometric albedo measurements taken simultaneously at BBSO, and
with effective albedo simulations for that night following Pall\'e
et al. (2003). To do that, we have averaged in wavelength each
spectrum between 480 and 700\,nm. Photometric observations from BBSO
cover the range between 400 and 700\,nm, including a major part of
the Rayleigh enhancement. Our simulations, however, cover the entire
range of short-wavelength radiation including ultraviolet, visible
and near-infrared (spanning 350 to 1500 nm).

Photometric observations at BBSO on 2003/11/19 were taken under a
moderately hazy sky, which increases the light scattered by the
moonshine and causes a too noisy result for that night; however, on
the following night, when the lunar crescent was smaller and the sky
conditions clearer, the noise was substantially reduced. The
retrieved photometric albedos, the spectrally  averaged albedos, and
computed photometric models are shown in Figure~\ref{mean}. Both
sets of observations were linearly fitted showing a comparable increase
from beginning to end of the night.

We regard all of these results as being consistent because of the
similarity of the increasing trend and because the offsets are those expected
from the different wavelength coverages. In particular, the photometrically
observed albedo includes more of the blue and less of the red, which would seem
to account for that offset. Further, the simulations include much more of the
IR, which roughly puts more weight toward an albedo of 0.25.

\section{Conclusions}

In this paper, we have accurately determined the Earth's apparent, global
spectroscopic albedo for a single night of November 2003 as measured from
Palomar Observatory in California. We show its main spectral features derived
from observations of the earthshine spectrum in the visible region. Our results
are consistent with simultaneous photometric apparent albedo measurements from
Big Bear Solar Observatory covering the same sunlit region of Earth. They are
also consistent with our albedo models using cloud patterns for the same day of
the year, although for different year due to the, so far, unavailability of the
necessary cloud cover data. The comparison with synthetic models of the Earth's
spectrum that take into account a precise cloud cover pattern for the night, and
its influence over surface and ocean distribution, will be carried out in future
works.

Our measurements do not show any sign of vegetation
red-edge, at least for the night of 2003/11/19. If this 
result were confirmed with the analysis of subsequent earthshine 
spectra taken at different seasons, and the analysis of
simultaneous cloud maps, it would strong limit efforts to use the 
red-edge as a probe for Earth-like extra-solar planets. We emphasize,
however, that other features on the earthshine spectra may still 
provide a powerful probe.

Observations of the earthshine at different seasons are desirable
not only to study other indirect biological signatures, as the
seasonal change of global abundances of oxygen, water vapor, carbon
dioxide or methane, but also for a better understanding of the
influence of these species over the Earth's albedo and climate.

Results derived from future observations of the earthshine will be
invaluable in the search for terrestrial-like extra-solar planets,
in particular the search of planetary atmospheres in chemical
disequilibrium (as proposed by Hitchcock and Lovelock, 1967). The
simultaneous detection of molecular oxygen and methane, for
instance, remains a potential biosignature.

\acknowledgments

This research was supported in part by a grant from NASA
(NASA-NNG04GN09G). Spectroscopic earthshine observations have been
partially supported by the Dudley Observatory through the 2002
Ernest F. Fullam Award.

\clearpage

%% Use the figure environment and \plotone or \plottwo to include
%% figures and captions in your electronic submission.
%% To embed the sample graphics in
%% the file, uncomment the \plotone, \plottwo, and
%% \includegraphics commands
%%
%% If you need a layout that cannot be achieved with \plotone or
%% \plottwo, you can invoke the graphicx package directly with the
%% \includegraphics command or use \plotfiddle. For more information,
%% please see the tutorial on "Using Electronic Art with AASTeX" in the
%% documentation section at the AASTeX Web site,
%% http://www.journals.uchicago.edu/AAS/AASTeX.
%%
%% The examples below also include sample markup for submission of
%% supplemental electronic materials. As always, be sure to check
%% the instructions to authors for the journal you are submitting to
%% for specific submissions guidelines as they vary from
%% journal to journal.

%% This example uses \plotone to include an EPS file scaled to
%% 80% of its natural size with \epsscale. Its caption
%% has been written to indicate that additional figure parts will be
%% available in the electronic journal.


\begin{thebibliography}{}

\bibitem[Arnold et al., 2002]{Arnold2002} Arnold L., S. Gillet, O. Lardiere, P.
Riaud, and J. Schneider, A test for the search for life on extrasolar planets.
Looking for the terrestrial vegetation signature in the Earthshine spectrum,
{\it Astronomy \& Astrophysics}, 392, 231-237, 2002.

\bibitem[Culf., 1995]{Culf1995} Culf A.D., G. Fisch and M.G. Hodnett,
The albedo of Amazonian forest and ranch land, {\it Journal of Climate}, 8, 6,
1995.

\bibitem[Goode et al., 2001]{Goode2001} Goode, P.R., J. Qiu, V. Yurchyshyn, J.
Hickey, M.C. Chu, E. Kolbe, C.T. Brown, and S.E. Koonin, Earthshine
observations of the Earth's reflectance, {\it Geophys. Res. Lett.}, 28 (9),
1671-1674, 2001.

\bibitem[Hapke, 1971]{Hapke1971} Hapke, B., Physics and Astronomy of the Moon,
2nd edition, ed. Z. Kopal, Academic Press, New York, 155, 1971.

\bibitem[Hapke et al., 1993]{hapke1993} Hapke, B.W., R.M. Nelson, and W.D.
Smythe, The opposition effect of the moon - the
contribution of coherent backscatter, {\it Science}, 260, 509-511, 1993.

\bibitem[Hapke et al., 1998]{Hapke1998} Hapke, B., R. Nelson, and W. Smythe, The
opposition effect of the moon: coherent backscatter and shadow hiding, {\it
Icarus}, 133, 89-97, 1998.

\bibitem[Helfenstein et al., 1997]{Helf1997} Helfenstein, P., Veverka, J., and
Hillier, J., The lunar
opposition effect: A test of alternative models, {\it Icarus}, 128, 2-14, 1997.

\bibitem[Hitchcock \& Lovelock, 1967] {Hitch1967} Hitchcock D.R. and Lovelock,
J.E., Life detection by atmospheric analysis, {\it Icarus}, 7, 149-159, 1967.

\bibitem[Monta\~n\'es Rodr{{i}}guez et al., 2004]{Montanes2004} Monta\~n\'es
Rodr{{i}}guez P., E. Palle, P.R. Goode, J. Hickey, J. Qiu, V. Yurchyshyn, M-C
Chu, E. Kolbe,C.T. Brown, S.E. Koonin. Advances in Space Research,
doi:10.1016/j.asr.2003.01.028, April 2004


\bibitem[In2001]{IPCC} Intergovernmental Panel on Climate Change (IPCC), 2001,
{\it The Scientific Basis, Contribution of Working Group I to the Third
Assessment Report of the
Intergovernmental Panel on Climate Change (IPCC)}, J. T. Houghton, Y. Ding, D.J.
Griggs, M. Noguer, P. J. van der Linden and D. Xiaosu (Eds.), Cambridge
University Press, pp 944, 2001.

\bibitem[Palle et al., 2004]{Palle2004}  Palle E., P. Monta\~n\'es
Rodr{{i}}guez, P.R. Goode, J. Hickey, J. Qiu, V. Yurchyshyn, M-C Chu, E.
Kolbe,C.T. Brown, S.E. Koonin. Advances in Space Research,
doi:10.1016/j.asr.2003.01.028, April 2004

\bibitem[Palle et al., 2003]{Palle2003} Pall\'e, E., P.R. Goode, V. Yurchyshyn,
J. Qiu, J. Hickey, P. Monta\~n\'es Rodr{{i}}guez,
M.C. Chu, E. Kolbe, C.T. Brown, and S.E. Koonin, Earthshine and the Earth's
albedo II: Observations and simulations over three years, {\it J. Geophys.
Res.}, 108(D22), 4710, doi: 10.1029/2003JD003611, 2003.

\bibitem[Palle et al., 2004b]{Palle2004b} Pall\'e E., P.R. Goode, P.
Monta\~n\'es Rodriguez and S.E. Koonin. Changes in the Earth's reflectance over
the past two decades. {\it Science}, 304, 1299-1301, doi:10.1126/science. 1094070, 28 May 2004b.

\bibitem[Qiu et al., 2003]{Qiu2003} Qiu, J., P.R. Goode, E. Pall\'e, V.
Yurchyshyn, J. Hickey, P. Monta\~n\'es-Rodr{{i}}guez,
M.C. Chu, E. Kolbe, C.T. Brown, and S.E. Koonin, Earthshine and the Earth's
albedo I: Earthshine observations and measurements of the lunar phase function
for accurate measurements of the Earth's Bond albedo, {\it J. Geophys. Res.}, 108(D22), 4709, doi: 10.1029/2003JD003610, 2003.

\bibitem[Rossow et al., 2004]{Rossow2004} Rossow, W.B., A.W. Walker, L.C. Garder, Comparison of ISCCP and OTHER Cloud Amounts, {\it Journal of Climate} vol. 6, Issue 12, pp. 2394-2418, 1993.


\bibitem[Wendisch et al., 2004]{ Wendisch2004} Wendisch, M., P. Pilewskie, E.
J\"akel, S. Schmidt, J. Pommier, S. Howard, H.H. Jonsson, H. Guan, M.
Schr\"oder, B. Mayer, Airborne measurements of areal spectral surface albedo
over different sea and land surfaces, {\it Journal of Geophysical Research},
109, D08203, doi:10.1029/2003JD004392,  2004.

\bibitem [Woolf et al., 2002] {Woolf2002} Woolf N.J., P.S. Smith, W.A. Traub and
K.W. Jucks, The Spectrum of Earthshine a pale blue dot observed from the ground,
{\it The Astrophysical journal}, 574, 430-433, 2002.

\end{thebibliography}
\end{document}